\documentclass[lettersize,journal]{IEEEtran}
\usepackage{lineno} 
\usepackage{amsmath,amsfonts}
\usepackage{algorithm}
\usepackage{algpseudocode}
\usepackage{array}
\usepackage[caption=false,font=normalsize,labelfont=sf,textfont=sf]{subfig}
\usepackage{textcomp}
\usepackage{stfloats}
\usepackage{url}
\usepackage{soul}
\usepackage{color}
\usepackage{stfloats}
\usepackage{tabularx}
\usepackage{verbatim}
\usepackage{graphicx}
\usepackage{ragged2e}
\newcolumntype{Y}{>{\RaggedRight\arraybackslash}X}
\def\BibTeX{{\rm B\kern-.05em{\sc i\kern-.025em b}\kern-.08em
    T\kern-.1667em\lower.7ex\hbox{E}\kern-.125emX}}
\usepackage{balance}

\begin{document}

\title{Bio-inspired Microgrid Management based on Brain’s Sensorimotor Gating}

\author{Panos Papageorgiou, Anastasios E. Giannopoulos, and Sotirios T. Spantideas

\thanks{
P. Papageorgiou is with the University of Patras, Greece (e-mail: papageorgiou@ece.upatras.gr).

A. Giannopoulos and S. Spantideas are with the National and Kapodistrian University of Athens, Athens, Greece (e-mail: \{angianno, sospanti\}@uoa.gr).

\textit{Corresponding author: A. Giannopoulos}
}}

\markboth{OCTOBER~2025}%
{Papageorgiou, 
\MakeLowercase{\textit{(et al.)}: 
Bio-inspired Microgrid Management based on Brain’s Sensorimotor Gating}}

\maketitle

\begin{abstract}
Microgrids are emerging as key enablers of resilient, sustainable, and intelligent power systems, but they continue to face challenges in dynamic disturbance handling, protection coordination, and uncertainty. Recent efforts have explored Brain Emotional Learning (BEL) controllers as bio-inspired solutions for microgrid control. Building on this growing trajectory, this article introduces a new paradigm for Neuro-Microgrids, inspired by the brain’s sensorimotor gating mechanisms, specifically the Prepulse Inhibition (PPI) and Prepulse Facilitation (PPF). Sensorimotor gating offers a biological model for selectively suppressing or amplifying responses depending on contextual relevance. By mapping these principles onto the hierarchical control architecture of microgrids, we propose a Sensorimotor Gating-Inspired Neuro-Microgrid (SG-NMG) framework. In this architecture, PPI-like control decisions correspond to protective damping in primary and secondary management of microgrids, whereas PPF-like decisions correspond to adaptive amplification of corrective control actions. The framework is presented through analytical workflow design, neuro-circuitry analogies, and integration with machine learning methods. Finally, open challenges and research directions are outlined, including the mathematical modeling of gating, digital twin validation, and cross-disciplinary collaboration between neuroscience and industrial power systems. The resulting paradigm highlights sensorimotor gating as a promising framework for designing self-protective, adaptive, and resilient microgrids.
\end{abstract}


\section{Introduction}

\IEEEPARstart{I}{n} the push toward a cleaner, more sustainable energy future, microgrids are emerging as essential building blocks for the modern power system \cite{olivares2014trends}. These localized energy systems, comprising distributed generation (solar, wind, etc.), energy storage, controllable loads, and advanced control/communication, can operate connected to the main grid or autonomously in islanded mode. Their dual-mode capability offers both operational flexibility and enhanced resilience. For example, microgrids can maintain power for critical infrastructure (hospitals, emergency services, community resilience hubs) during extreme weather or grid outages, while in normal operation helping to balance local supply and demand, reduce transmission losses, and support renewable penetration \cite{bacha2015photovoltaics}.

Beyond resilience, microgrids play a central role in achieving sustainability and energy independence. As policies and market mechanisms worldwide increasingly favor decarbonization and electrification, microgrids enable high integration of intermittent renewable energy sources (RES), storage systems, smart loads, and demand response strategies \cite{etxeberria2010hybrid}. They also facilitate energy autonomy for remote or underserved communities, reduce exposure to central grid failures, and allow experimentation with innovative control architectures and business models. Together, these attributes make microgrids crucial for future smart, resilient, and sustainable energy systems.

Despite their promise, microgrids face serious challenges in design and operation, especially under dynamic and uncertain conditions \cite{saeed2021review}. Their operation must deal with frequent disturbances such as sudden load changes, renewable generation variability, fault conditions, switching transients, and islanding transitions. Ensuring voltage and frequency stability across multiple timescales becomes more complex as the number of distributed resources increases and traditional system (slow) responses are reduced. Protection coordination is also more difficult due to bidirectional power flows, non‐standard fault currents, and changing network topology complicate conventional protection schemes. Finally, system uncertainty (in terms of RES output, load demand, or system conditions) and the requirement to provide robust protection (avoiding both false trips and missed faults) demand adaptive, fast, and reliable control architectures \cite{altaf2022microgrid}.

\subsection{Imitating Human Brain for Microgrid Design}

The brain has long served as a rich source of inspiration for control systems \cite{shadmehr2023biological}. Brain-inspired design promises several advantages such as fast reflexive reactions to threats, adaptability through learning, robustness under uncertainty, and ability to generalize from experience \cite{saavedra2024brain, yeganeh2022intelligent,li2024provisional,gu2024brain}. In microgrids, where disturbances, component failures, or nonlinearities are frequent, adopting architectures that mimic the brain’s multi-layered or hierarchical control and learning can improve response time, reduce overshoot, prevent false alarms, and maintain stability while preserving flexibility \cite{panda2022review,de2024brain}. Thus, brain-inspired controllers like Brain Emotional Learning (BEL) systems have been recently introduced, offering paths to react fast (as a reflex), adapt (through learning), and regulate (through higher-level inhibition) in ways conventional Proportional–Integral–Derivative (PID) controllers or lookup-table methods cannot ensure efficiently \cite{gu2024brain,mandal2024bioinspired}.

There are already several works that apply brain-inspired or BEL methods to microgrid control and protection \cite{jha2022online}. For example, the authors in \cite{li2024provisional} proposed a BEL-based Intelligent Controller (BELBIC) for frequency regulation in a hybrid AC/DC microgrid and implemented it on Field Programmable Gate Array (FPGA), demonstrating robustness under load changes and uncertainties. Another paper proposed a BELBIC controller for islanded microgrid frequency regulation and compared it against PID, Particle Swarm Optimization (PSO)-PID, Teaching-Learning-based Optimization (TLBO), and Grey wolf optimizer (GWO) controllers, showing reduced overshoot and faster settling times under disturbances \cite{gu2024brain}. Other works have highlighted multiple brain modeling works for microgrid management, mapping the emotional learning paradigm (including amygdala, orbitofrontal cortex, thalamus, and sensory cortex) to hierarchical control in microgrids \cite{saavedra2024brain,de2024brain}. Also, another study developed a secondary control law inspired by emotional learning in mammalian limbic systems to handle uncertainties and faults with good performance under simulation \cite{khalghani2016self}. More broadly, emerging research on Artificial Intelligence/Machine Learning (AI/ML)-based microgrid hierarchical control (e.g. secondary/tertiary layers) often mention or include BEL approaches among the set of techniques for adaptive control and protection \cite{li2023hierarchical}.


\subsection{Objectives and Contributions}

Building on the growing application of brain-inspired approaches, this article describes and proposes an additional brain modeling paradigm, which is originated by the brain’s sensorimotor gating \cite{swerdlow2016sensorimotor}. Sensorimotor gating involves a network of synergistic neural circuits that act as a fundamental filtering system, determining which (and to what extent) sensory inputs trigger responses and which are suppressed \cite{franklin2011computational}. Specifically, similar to the wide use of BEL modeling, Prepulse Inhibition (PPI) and Prepulse Facilitation (PPF) brain mechanisms \cite{giannopoulos2024prepulse}, which are strongly coupled with the sensorimotor gating function, are envisioned as promising paradigms for guiding the development and testing of a new microgrid controller family, with the goal of advancing both control and protection capabilities. Imitating the brain's sensorimotor gating, the target is to leverage PPI/PPF as mechanisms for proactive microgrid disturbance response to detect precursor events or “prepulses” (e.g., small transient disturbances), and decide whether to suppress or boost subsequent control or protection responses depending on the profiling (e.g., timing and context) of early system failures. This is able to allow the system to avoid overreacting to benign fluctuations (false positives) and act dynamically when early warning signs predict serious upcoming disturbance. To this end, we present and map the neuronal computational model of PPI/PPF response loop (including brain circuitry like thalamus, sensory cortex, amygdala, prefrontal cortex) to a Neuro-Microgrid (NMG) hierarchical control architecture for fast reflex protection, containing NMG system components like analytics, supervisory inhibition/facilitation, gate logic, and actuators. Finally, a generalized hierarchical sensorimotor gating-inspired NMG controller architecture is outlined, supporting the decoupling of primary and secondary control loops. We propose that embedding this gating mechanism into microgrid secondary control layers can be a promising architecture for better disturbance handling, fewer unnecessary protection actions, and more resilient operation.

Based on the above, the main contributions of this article are fourfold. First, we introduce the neuroscience background of sensorimotor gating and the respective computational model workflow, with emphasis on PPI and PPF mechanisms. To this end, we describe the brain circuits underlying sensorimotor responses and their functional role in filtering or amplifying incoming sensory signals. Second, we draw explicit analogies between hierarchical microgrid control (primary, secondary, tertiary) and the gating functions of the brain. Third, we propose a conceptual Neuro-Microgrid architecture in which sensorimotor gating principles regulate control and protection layers, enabling proactive disturbance management. The conceptual and functional principles of the proposed architecture are outlined in detail, describing how the control/workflow steps evolve and how the suppression of nuisance events is achieved while facilitating rapid action against harmful disturbances. Fourth, we highlight open challenges and research opportunities, including mathematical modeling of gating factors, timing coordination across microgrid control layers, real-time implementation, and validation. Overall, the paper contributes an interdisciplinary perspective by bridging neuroscience insights with power system engineering, thereby extending the scope of brain-inspired microgrid research beyond emotional learning toward new paradigms of gating-based adaptive control.



\section{Sensorimotor Gating in the Brain}

\subsection{Sensorimotor Gating: The Brain's Automatic Filter}

Sensorimotor gating is a fundamental filtering and regulation process of the nervous system that prevents sensory overload by reducing the flow of irrelevant or redundant information to higher cortical centers. In essence, it allows organisms to maintain adaptive responses to important stimuli while ignoring trivial background noise, thereby preserving attentional resources and ensuring efficient behavioral control \cite{swerdlow2016sensorimotor,golubic2019attention}. In simplistic words, sensorimotor gating system acts as a gatekeeper to protect the brain from sensory overload and process efficiently incoming stimuli, blocking unnecessary information and adjusting the extent to which sensory inputs are processed \cite{velasques2011sensorimotor}. Without such gating, every stimulus would compete for processing, leading to cognitive fragmentation and impaired behavioral performance. This mechanism is crucial for everyday functioning, and its disruption has been linked to neurological and psychiatric conditions such as schizophrenia, Huntington’s disease, and Parkinson’s disease \cite{powell2002developmental, giannopoulos2021instantaneous,braff1990sensorimotor}.

The neurobiological circuitry of sensorimotor gating spans multiple hierarchical levels of the brain \cite{egger2020neural}. Sensory information is first relayed via the thalamus to both subcortical and cortical structures. The sensory cortex generates refined stimulus representations, which are integrated by the amygdala and striatum to evaluate salience and guide adaptive responses. Importantly, the prefrontal cortex exerts top-down inhibitory control, ensuring that exaggerated or inappropriate responses are suppressed, while the pontine reticular nucleus (PnC) in the brainstem acts as the final motor gateway for startle and reflex behaviors \cite{schouenborg2004learning}. This distributed loop allows rapid reflexive responses to salient events while maintaining flexibility through cortical modulation. In this way, sensorimotor gating serves as a dynamic filter–regulator system, balancing fast survival mechanisms (primary control) with higher-order supervisory functions (secondary/tertiary control) in the brain.

\subsection{Prepulse Inhibition as Protective Damping}

\begin{figure}[!htbp]
\centering
\includegraphics[trim={0cm 1cm 13cm 26cm},clip,width=1\columnwidth]{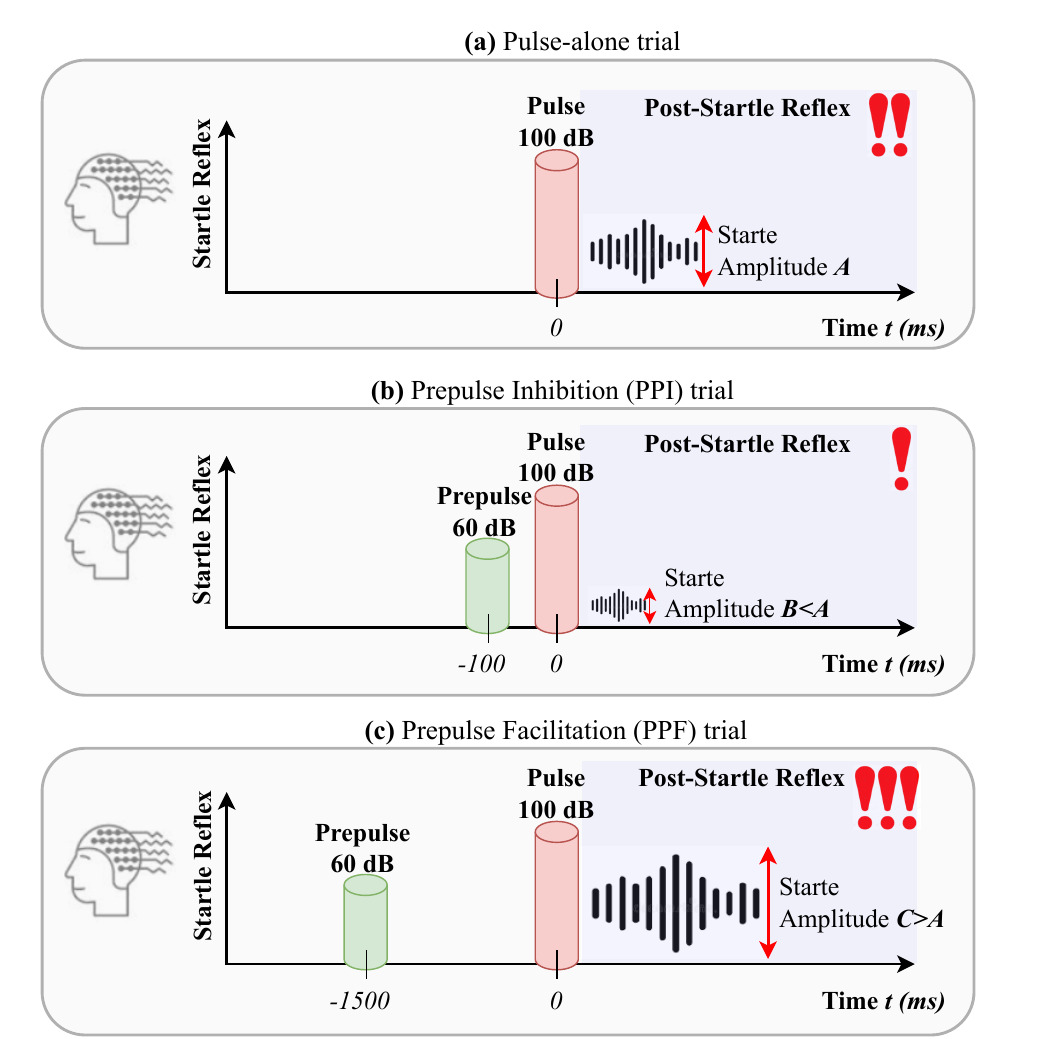}
\caption{Measuring the performance of Sensorimotor Gating via Startle Reflex. \textbf{(a)} Pulse-alone single-trial recording with startle reflex $A$; \textbf{(b)} PPI single-trial recording with startle reflex $B<A$ (Inhibition); \textbf{(c)} PPF single-trial recording with startle reflex $C>A$ (Facilitation).}
\label{fig:fig1}
\end{figure}

The most common metric of quantifying the performance of sensorimotor gating is the startle reflex \cite{giannopoulos2022evaluating}. The startle reflex is a rapid, automatic defensive reaction to sudden and intense stimuli, such as a loud sound or a bright flash, as depicted in Fig.~\ref{fig:fig1}(a). It is mediated primarily at the brainstem level, through the caudal pontine reticular nucleus (PnC), which relays to spinal and cranial motor neurons \cite{ramirez2012computational}. In humans, the startle reflex is often measured as an eye-blink response (orbicularis oculi muscle contraction), though it can also manifest as whole-body jerks or muscle contractions \cite{maslovat2021response}. In experimental paradigms, startle amplitude is quantified in multiple pulse-alone trials, where a strong stimulus is presented in isolation.

Startle reflex can be altered and modulated depending on the incoming stimuli. Two of the most-studied operational measures of sensorimotor gating are the PPI and PPF, which both lead to the modulation of startle reflex \cite{cano2021amygdala,giannopoulos2022evaluating}. PPI refers to the reduction in the amplitude of a startle reflex when a weak, non-startling stimulus (the “prepulse”) precedes a stronger startling stimulus (the “pulse”) by a short interval, typically between 30 and 500 ms, as depicted in Fig.~\ref{fig:fig1}(b). The prepulse does not itself trigger a startle response, but it primes the nervous system to filter or gate the subsequent stimulus, resulting in an attenuated behavioral response \cite{giannopoulos2022evaluating}. PPI is considered “pre-attentive” that it occurs automatically, though higher-level cognitive factors like attention, emotion, or expectancy can modulate its magnitude. Researchers use PPI as a metric of how well the nervous system gates irrelevant sudden sensory inputs. In this sense, the primary role of PPI is to avoid sensory overload by preventing irrelevant or redundant stimuli from eliciting full reflexive reactions. This enhances the robustness of neural responses, allowing the organism to conserve resources and maintain focus on relevant signals \cite{swerdlow2016sensorimotor}. Clinically, PPI is used as a biomarker of sensorimotor gating efficiency, with deficits linked to conditions such as schizophrenia, Tourette’s syndrome, and Huntington’s disease \cite{santos2023prepulse}.

The circuitry underlying PPI involves both subcortical reflex arcs and top-down cortical modulation. The thalamus acts as the first relay of sensory information, sending rapid input to the brainstem startle circuit (particularly the caudal Pontine Reticular Nucleus, PnC). When a prepulse is present, cortical and limbic structures modulate this pathway. The Prefrontal Cortex (PFC) provides inhibitory control over the amygdala and striatal circuits, reducing excitatory drive to the PnC. The striatum and hippocampus contribute context and memory signals that shape gating efficiency \cite{cano2021amygdala,ramirez2012computational}. As a result, the prepulse engages a distributed network that suppresses brainstem excitability, thereby reducing startle magnitude.

\subsection{Prepulse Facilitation as Adaptive Amplification}

In contrast to the PPI, PPF refers to the enhancement of a startle response when a weak prepulse precedes a startling stimulus under specific temporal or contextual conditions. Unlike PPI, which suppresses responses, PPF increases the magnitude of the reflex (see Fig.~\ref{fig:fig1}(c)), typically observed when the interval between prepulse and pulse is longer (e.g., greater than 500 ms) or when attentional/arousal states bias the system toward heightened responsiveness \cite{naysmith2021neural}. Thus, PPF tends to occur when the prepulse-pulse interval is longer, or when the nervous system is in a state of heightened arousal or attention \cite{giannopoulos2022evaluating}. The functional role of PPF is to boost sensitivity to relevant signals, especially in conditions where detecting weak but behaviorally significant events is critical. PPF can be interpreted as an adaptive amplification mechanism, ensuring that organisms remain vigilant and responsive in contexts where environmental cues predict important outcomes.

The neural basis of PPF overlaps with that of PPI but shifts the balance toward excitatory facilitation rather than inhibition \cite{cano2021amygdala}. Prepulses engage the thalamus and sensory cortex, which provide detailed stimulus representations. The hippocampus and striatum encode contextual predictions and temporal expectations, while the PFC modulates excitatory gain rather than inhibition. This increases drive from cortical–limbic circuits to the brainstem PnC, thereby enhancing the startle reflex \cite{bezerra2024computational}. PPF thus represents the other side of the gating spectrum, transforming a prepulse from an inhibitory signal into a facilitatory one under the right conditions.

\subsection{Relevance and Mapping to System Control}

From a control theory perspective, the mechanisms of PPI and PPF can be understood as two complementary modes of dynamic regulation. PPI functions as a protective damping mechanism, suppressing excessive or unnecessary responses to non-critical events. By transiently lowering the excitability of the startle circuit, PPI ensures that sudden but irrelevant inputs do not destabilize the system, thus maintaining stability and preventing over-reaction \cite{swerdlow2016sensorimotor}. This is analogous to damping in engineered control systems, where high-frequency or low-significance disturbances are attenuated to protect the system from instability. In contrast, PPF acts as an adaptive amplification mechanism, selectively boosting responsiveness when preparatory cues signal that an important event is imminent. By increasing excitatory drive under specific timing and contextual conditions, PPF ensures that the system becomes more sensitive and reactive to relevant stimuli \cite{miller2021robust}. This is conceptually similar to gain scheduling or adaptive control in engineering, where system sensitivity is strategically increased to capture weak but critical signals.

Together, PPI and PPF embody a dual-mode gating strategy that balances stability and sensitivity. This balance is essential for robust operation, which means that the system remains resistant to nuisance inputs yet agile enough to amplify early warnings of significant events. Translating this principle to engineered systems, such as microgrids, suggests that adaptive gating mechanisms could help achieve a similar balance between protective stability (avoiding unnecessary trips) and proactive responsiveness (amplifying preventive actions against faults). At a higher level, a conceptual mapping between PPI/PPF and microgrid hierarchical control can be established:

\begin{enumerate}
    \item The Prepulse (small disturbance) may be mapped to a precursor event (e.g., voltage sag, transient harmonic, short load fluctuation).
    \item The Pulse (startle stimulus) can represent the major event/fault (e.g., short-circuit, severe imbalance, islanding).
    \item PPI (inhibition) mechanism corresponds to protective damping in primary/secondary control, which means that the nuisance transients are filtered, avoiding unnecessary breaker trips or control oscillations.
    \item PPF (facilitation) mechanism corresponds to adaptive amplification in secondary/tertiary control. This means that precursor events are interpreted as early warnings, boosting corrective actions such as fast energy storage injection or adaptive setpoint tuning.
    \item The brain interaction loop across Thalamus (Th), Sensory Cortex (SC), Prefrontal Cortex (PFC), Amygdala (AMYG) and Striatum loop may be mapped to a closed control loop including Microgrid Sensors, Analytics module, Energy Management System (EMS), Reflex Controller and Gate Logic modules in NMGs, where EMS plays the role of cortical supervision deciding whether to inhibit or facilitate reflexive protection.
\end{enumerate}

\section{Neuro-Microgrids: Neuro-Inspired Microgrid Control}

\subsection{Microgrid Components}

\begin{figure*}[!htbp]
\centering
\includegraphics[trim={0cm 0cm 8cm 11cm},clip,width=2\textwidth]{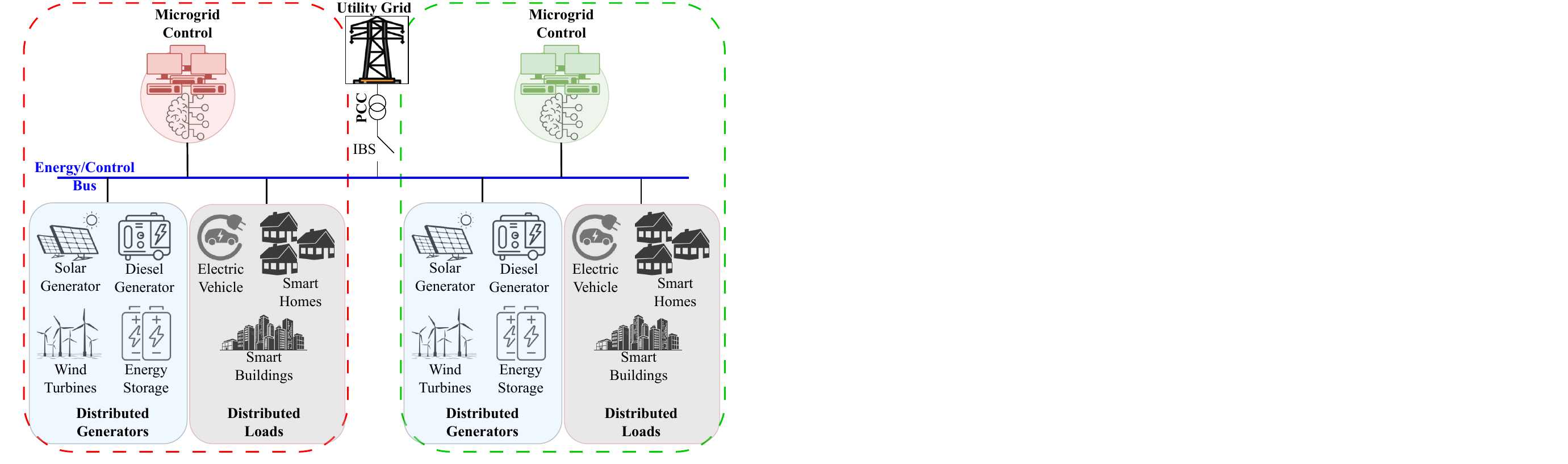}
\caption{High-level Multi-Microgrid Architecture. PCC: Point of Common Coupling; IBS: Intelligent Bypass Switch}
\label{fig:fig2}
\end{figure*}

As depicted in Fig.~\ref{fig:fig2} a microgrid is a localized cluster of energy resources and loads, designed to operate both in connection with the main grid and in islanded mode \cite{farrokhabadi2019microgrid}. Its main components include distributed generation (DG), energy storage systems (ESS), and controllable or distributed loads (DLs). DG units encompass renewable sources such as photovoltaic panels and wind turbines, as well as dispatchable generators like microturbines and fuel cells \cite{vasquez2010hierarchical}. ESS, such as batteries or supercapacitors, provide balancing services by absorbing excess generation and supplying power during shortages, thereby smoothing renewable variability and enhancing resilience \cite{zhou2016review}. Controllable loads, including flexible industrial processes, Heating-Ventilation-Air Conditioning (HVAC) systems, or electric vehicle (EV) charging, enable demand response by adjusting consumption according to grid conditions or economic signals \cite{guerrero2012advanced}. Together, these elements form a cyber-physical system where monitoring, communication, and control are as critical as the physical assets.

When grid-connected, microgrids interact dynamically with the utility grid to exchange power through the Point of Common Coupling (PCC), participate in demand response programs, and contribute ancillary services such as frequency regulation or reactive power support. In this mode, the Microgrid Management System (MGMS) must optimize energy flow between local resources and the grid, often prioritizing economic and market signals while maintaining local stability \cite{guerrero2012advanced}. By contrast, when islanded, the microgrid must operate autonomously, relying solely on its internal DGs, ESS, and load management to maintain frequency and voltage stability. Here, the MGMS takes on a more protective and stabilizing role, ensuring that critical loads are supplied, non-critical loads may be shed, and resources are dispatched adaptively to counteract disturbances and uncertainty \cite{guerrero2010hierarchical}. Thus, in both modes, controllers play a pivotal role in orchestrating distributed resources, switching seamlessly between economic optimization in grid-connected mode and survival-oriented stabilization in islanded mode.

\subsection{Hierarchical Control Architecture}

The operation of microgrids is based on the Hierarchical Control Architecture (HCA), typically structured under a three-layer hierarchical control framework \cite{guerrero2012advanced}. Primary control operates at the fastest timescale (milliseconds to seconds) and is responsible for maintaining local voltage and frequency stability. Droop control methods are commonly employed, allowing distributed generators to share load proportionally without relying on high-bandwidth communication \cite{guerrero2010hierarchical}. Secondary control acts at a slower timescale (seconds to minutes), correcting the residual deviations left by primary control. Its role is to restore system variables (e.g., frequency, voltage) to nominal values and to coordinate the operation of multiple units, often through centralized or distributed algorithms \cite{farrokhabadi2019microgrid}. Tertiary control governs the longest timescale (minutes to hours) and addresses economic and strategic objectives. It performs optimization, scheduling, and market participation, deciding when to draw from or inject power into the main grid, how to dispatch distributed resources, and how to maximize both technical efficiency and economic benefit \cite{katiraei2006power,spantideas2025autonomous}.

Overall, this layered architecture ensures both stability and flexibility, meaning that primary control guarantees survival, secondary provides coordination, and tertiary aligns microgrid operation with broader system and market goals. Yet, as disturbances grow more frequent and uncertainty increases, traditional hierarchical control faces limitations, motivating exploration of neuro-inspired strategies that can emulate the brain’s ability to combine reflexive action, adaptive correction, and long-term planning.

\subsection{Existing Brain-inspired Designs}

\subsubsection{Brain Emotional Learning-based Intelligent Controllers}

The widely-used BEL/BELBIC controllers abstract the limbic circuit into four cooperating modules, namely the Thalamus (Th) and Sensory Cortex (SC) as sensory relays/feature extractors, the Amygdala (AMYG) as a fast excitatory pathway that predicts reinforcement (drives an emotional output), and the Orbitofrontal Cortex (OFC) as an inhibitory supervisor that suppresses inappropriate/excessive responses \cite{saavedra2024brain}. In control terms, Th/SC transform raw measurements into features, AMYG produces a fast corrective drive, and OFC compares expected vs. actual outcomes and inhibits the AMYG when the controller over-reacts \cite{gu2024brain}. Canonical BEL diagrams show the dual pathways (Th-AMYG and Th-SC-AMYG/OFC) with an inhibitory link OFC-AMYG, and learning rules driven by a reinforcement signal (truth/utility) that splits the prediction error so AMYG learns when it under-predicts and OFC learns when AMYG over-predicts \cite{saavedra2024brain,de2024brain}. This yields a reflex-plus-supervision loop that is robust to model errors and nonlinearity and is implementable in secondary control timescales.

\begin{figure*}[!htbp]
\centering
\includegraphics[trim={0cm 1cm 12cm 26cm},clip,width=2\textwidth]{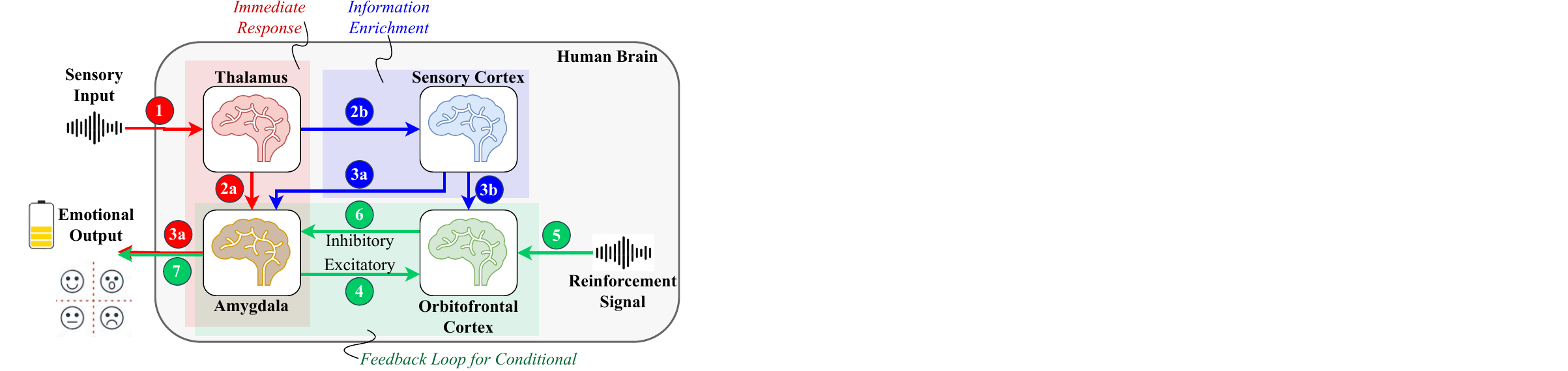}
\caption{General Neuro-circuitry and Workflow of Brain Emotional Learning Controllers.}
\label{fig:fig3}
\end{figure*}

The current principles underlying the design of BEL-based controllers are rooted in simplified brain computational models of the limbic system, where sensory input, emotional evaluation, and inhibitory supervision interact to generate adaptive responses. As illustrated in Fig.~\ref{fig:fig3}, the BEL architecture includes four core modules, including Th (which rapidly relays crude sensory signals), SC (which provides more refined feature representations), AMYG (which computes a fast excitatory output based on stimulus-reinforcement signals associations), and OFC (which learns inhibitory responses that regulate the AMYG when it overestimates the required reaction.) In computational terms, the AMYG produces an initial control signal driven by both Th and SC inputs, while the OFC supplies an inhibitory correction term. Both structures update their synaptic weights using a reinforcement signal, which serves as the “ground truth” of system performance \cite{saavedra2024brain}.

The generalized workflow of BEL circuitry mirrors this biological organization, as shown in Fig.~\ref{fig:fig3}. A sensory input (stimulus) enters through the Thalamus, which immediately sends a signal to the AMYG for a rapid and reflex-like corrective response (path of steps 1-2a-2b), and in parallel to the SC for refined processing (step 2b). The SC output reaches both AMYG and OFC, providing contextually-enriched information for evaluating the stimulus (steps 3a and 3b). The AMYG then generates an excitatory drive, which constitutes the preliminary control action (step 4). At the same time, the OFC processes the SC input and compares it against reinforcement feedback (step 5): if the AMYG’s action is too strong relative to the desired response, the OFC sends an inhibitory signal to attenuate AMYG activity (step 6). The final BEL controller output is thus the difference between excitatory and inhibitory pathways (step 7), ensuring fast yet regulated responses. In this sense, pathway 4-5-6-7 constitutes a feedback loop for conditional correction of the initial AMYG output. Over time, the AMYG learns to predict the reinforcement outcome for relevant stimuli, while the OFC improves its capacity to inhibit over-reactions, yielding a balance between responsiveness and stability.

To better illustrate the operation of BEL from the brain perspective, consider the case of a sudden loud noise as the sensory input. The Thalamus quickly relays a crude signal to the Amygdala, which instinctively prepares a strong startle or defensive reaction. In parallel, the SC processes the audiovisual features and identifies the source more precisely. For instance, SC can enrich the raw signal with information about whether the sound came from a book falling or from a genuinely threatening event such as an intruder. If the noise is harmless, the OFC, informed by cortical processing and reinforcement feedback from past experiences, sends an inhibitory signal to dampen the AMYG’s overreaction, reducing the behavioral output. Conversely, if the noise is judged threatening, the OFC allows the AMYG response to proceed with minimal inhibition, resulting in a full defensive reaction. In this way, BEL circuitry ensures that the organism remains fast and reflexive, yet avoids wasting energy on inappropriate responses to benign stimuli.

To map this mechanism in microgrid terms, consider a microgrid frequency control task. A sudden load increase acts as the disturbance. The Thalamus module relays this event directly to the AMYG, which produces a strong initial correction by requesting higher power injection at the DG. Simultaneously, the SC processes the magnitude and persistence of the disturbance, sending contextual information to both AMYG and OFC. The AMYG tends to overshoot by demanding more correction than necessary, but reinforcement feedback indicates that such an overshoot destabilizes voltage. The OFC then sends an inhibitory signal to scale back the AMYG output, resulting in a final control action that balances rapid frequency restoration with voltage stability. In this way, the BEL controller combines the speed of reflexive response with the safeguards of supervisory inhibition, a principle directly inspired by limbic neuro-circuitry.

A series of papers apply BELBIC to frequency regulation in islanded or provisional (hybrid AC/DC) microgrids with renewables and uncertain loads \cite{yeganeh2022intelligent,lucas2004introducing}. These works consistently report that BEL-based controllers offer reduced overshoot and settling time compared with classical/optimized PI/PID under disturbances and RES variability, with some demonstrating embedded realizations (e.g., FPGA) for practical fast loops \cite{li2024provisional}. Examples include (i) BELBIC for islanded microgrid frequency control with comparative evaluation versus PID/PSO-PID/TLBO/GWO designs \cite{gu2024brain}, (ii) BELBIC frequency control of provisional MGs \cite{gu2024belbic}, and (iii) an AC/DC hybrid implementation emphasizing robustness across operating modes \cite{albalawi2022analysis}. Other studies have synthesized this trajectory and explicitly map Th–SC–AMYG–OFC to primary/secondary/tertiary layers in MGs, highlighting BEL benefit as an adaptive secondary controller that corrects residual deviations left by primary droop \cite{saavedra2024brain}. BEL/BELBIC has also been explored in adjacent industrial electrical domains (e.g., flow/pumping systems) as a drop-in intelligent controller, further evidencing generalizability and compatibility with heuristic/metaheuristic tuning \cite{marques2022emotional}. 

\subsubsection{Reinforcement Learning and Neuro-fuzzy Methods}

Reinforcement learning (RL) and Deep RL (DRL) have been used for secondary control and energy management, addressing nonlinearity and uncertainty with learning-based policies. For instance, event-triggered RL has been proposed to reduce communication burden while restoring voltage/frequency in DC microgrids \cite{negahdar2024reinforcement} and AC islanded microgrid \cite{li2024event}. Another study has used DRL for real-time economic energy management \cite{liu2023deep}, while other works have developed distributed/stochastic DRL for islanded AC microgrids integrating frequency/voltage restoration with proportional sharing \cite{dehkordi2025adaptive}. There have been also studies that summarize strengths (coordination, long-horizon optimization) and barriers (sample efficiency, safety constraints in protection-grade loops) of RL and AI in microgrids \cite{barbalho2025reinforcement,waghmare2025systematic}. In parallel, fuzzy and neuro-fuzzy approaches remain attractive for uncertainty handling without full models, including type-1/type-2 fuzzy methods \cite{khooban2020novel}, Adaptive Neuro-Fuzzy Inference System (ANFIS) \cite{karkevandi2018anfis}, and neural network-assisted adaptive methods for Proportional-Integral/Proportional-Resonant (PI/PR) controllers \cite{liu2022neural,sun2017artificial}. All these methods have shown improved transient tracking and disturbance rejection in inverter-dominated microgrids and small isolated systems. Collectively, these algorithmic families give us reactive adaptation (post-event correction), robustness to uncertainty, and practical implementability across secondary/tertiary layers. Table~\ref{tab2} summarizes some of the key related works about different domains of brain-inspired microgrid design.

\begin{table*}[!htbp]
\caption{Brain-inspired and AI-based Control Techniques for Microgrids}
\label{tab2}
\centering
\begin{tabularx}{\textwidth}{|Y|Y|Y|Y|Y|}
\hline
\textbf{Controller Family} & \textbf{Related Works} & \textbf{Target Layer for Control} & \textbf{Key Contributions} & \textbf{Sensorimotor Gating Analogy}\\
\hline
BEL/BELBIC (Brain Emotional Learning) & \cite{li2024provisional,gu2024belbic,saavedra2024brain,lucas2004introducing,albalawi2022analysis,yeganeh2022intelligent} & Secondary (frequency \& voltage restoration); sometimes hybrid AC/DC & Brain-inspired Th–SC–AMYG–OFC workflow; fast excitatory + inhibitory learning; reduced overshoot and settling; FPGA feasibility & Inhibition via OFC present, but no explicit conditioning on precursor–disturbance timing\\
\hline
Reinforcement Learning (RL/DRL) & \cite{negahdar2024reinforcement,li2024event, liu2023deep,dehkordi2025adaptive, barbalho2025reinforcement,waghmare2025systematic} & Secondary (restoration, coordination); Tertiary (economic optimization) & Learns optimal policies under uncertainty; coordination across agents; long-horizon optimization; adaptive energy scheduling & Policy adaptation reactive to states/rewards; lacks fast gating of nuisance vs. precursor events\\
\hline
Fuzzy \& Neuro-Fuzzy Controllers & \cite{khooban2020novel,karkevandi2018anfis} & Secondary (voltage–frequency regulation); sometimes Primary support & Rule-based adaptation; robust under model uncertainty; improved transient tracking; heuristic/metaheuristic tuning possible & No explicit temporal gating; heuristic adaptation post-disturbance\\
\hline
Artificial Neural Networks (ANN/Hybrid) & \cite{liu2022neural,sun2017artificial} & Secondary (tracking); Tertiary (forecasting/management) & Adaptive learning of nonlinear dynamics; good at prediction and optimization tasks & No explicit suppression or facilitation mechanisms; lacks context-dependent timing filters\\
\hline
\end{tabularx}
\end{table*}

\subsection{The Gaps of Existing Brain-inspired Microgrid Controllers}

Despite progress, current approaches rarely implement a conditional (e.g., timing-aware), dynamic “gating” mechanism that (i) suppresses nuisance disturbances before they cascade (protective damping) and (ii) proactively amplifies control readiness when precursor signatures predict serious events (adaptive facilitation). BEL/BELBIC provides fast correction with supervisory inhibition, yet it typically learns from realized deviations rather than conditioning responses on precursor-disturbance features (e.g., timing or historical patterns). RL/DRL excels at long-horizon optimization and coordination, but safety-critical pre-emptive filtering of short-timescale precursors is not explicit, and training stability/safety constraints remain challenging in real-time protection loops. Fuzzy/neuro-fuzzy controllers capture heuristic adaptation but usually lack an explicit temporal gating factor that modulates action gain as a function of inter-event interval and context. Meanwhile, studies on microgrid protection continue to emphasize difficulties with bidirectional flows, non-standard fault currents, and topology changes, where false trips and missed faults remain a central trade-off. Consequently, BEL/BELBIC methods provide the closest analogy to neuro-circuits but only covers reactive inhibition/supervision, not conditional gating. The same stands for fuzzy/ANN methods that provide robustness, but they are largely heuristic and reactive. As a result, none of the existing methods explicitly implement gating functions analogous to sensorimotor filtering system. This may further motivate importing sensorimotor gating from neuroscience into the hierarchical control stack so the microgrid can take advantage of early warnings to proactively handle upcoming failures.

\section{Parallels Between Sensorimotor Gating and Neuro-Microgrids}

\begin{figure*}[!htbp]
\centering
\includegraphics[trim={0cm 0cm 10cm 10cm},clip,width=1\textwidth]{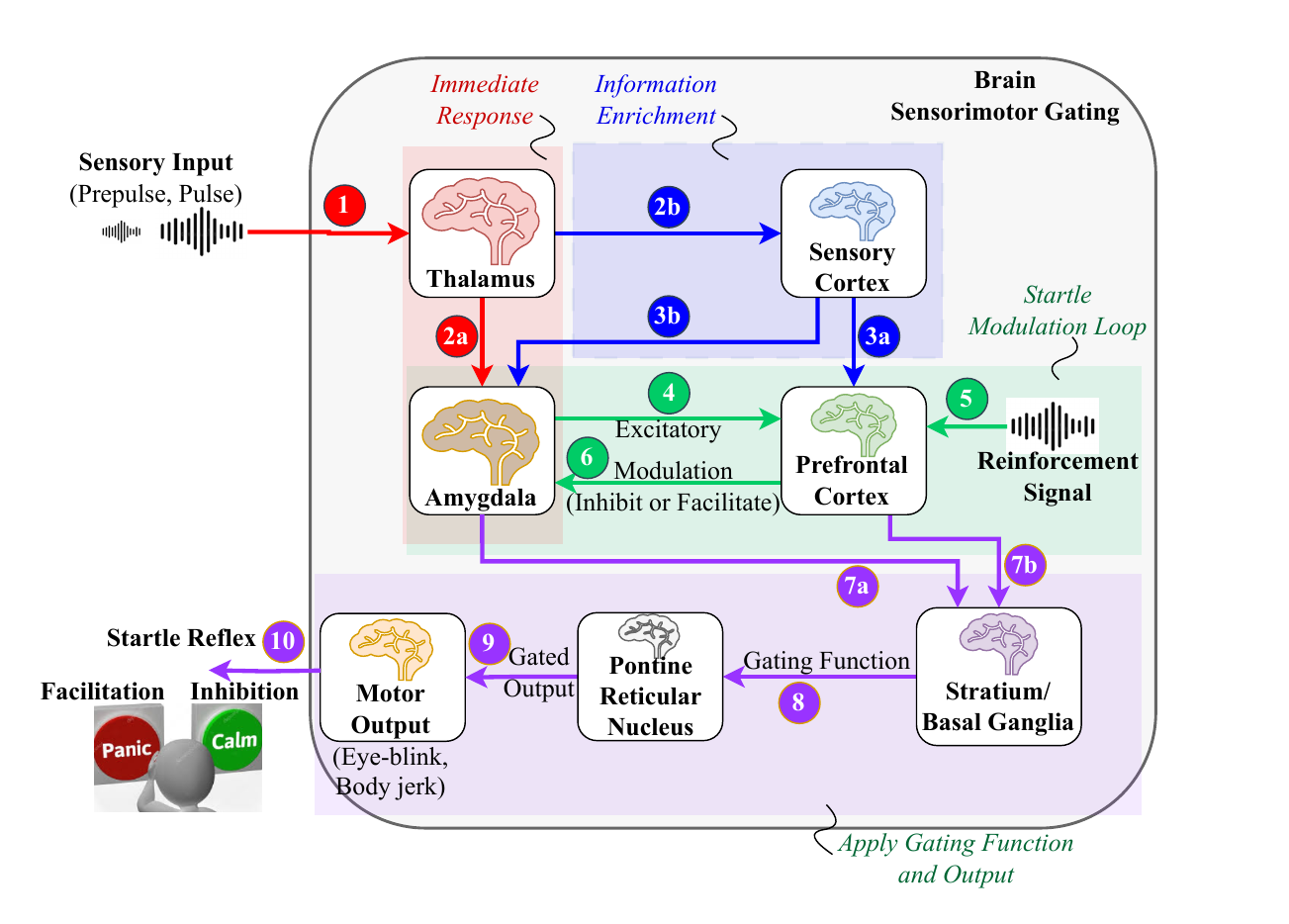}
\caption{Simplified Neuro-circuitry and Workflow of Sensorimotor Gating Pathways. A prepulse sets a timing-dependent gating factor via cortical–basal ganglia control over the PnC startle gateway, producing inhibition (PPI) or facilitation (PPF) of the subsequent startle output. AMYG–PFC act as salience and inhibitory supervisors analogous to BEL models.}
\label{fig:fig4}
\end{figure*}

In Fig.~\ref{fig:fig4}, we illustrate the simplified neuro-circuitry and workflow (or computational model) of sensorimotor gating in the human brain \cite{franklin2011computational,ramirez2012computational,bezerra2024computational}. Sensory inputs (prepulse and pulse) are first relayed by the Thalamus, which provides a fast but crude representation to the AMYG (immediate response) while simultaneously passing the signal to the SC for enrichment. The SC refines stimulus features and communicates them to both the AMYG and the PFC. The AMYG then generates a rapid excitatory modulation, essentially a reflexive tendency to react (let's call it $A(t)$), while the PFC supplies an inhibitory or facilitatory modulation (let's call it $I_{PFC \rightarrow AMYG}$) informed by contextual knowledge and reinforcement feedback. This interplay determines whether the reflexive action is suppressed (PPI) or amplified (PPF). Downstream, the Striatum/Basal Ganglia integrates these signals and applies a gating function $g(\Delta t)$ ($\Delta t$ is the Prepulse-Pulse interval) to the Pontine Reticular Nucleus (PnC), which is the motor gateway for startle reflexes. The final motor output, measured as eye-blink or body jerk, is thus the gated result of excitatory drive, inhibitory control, and timing-based gating modulation. In this sense, the final output can be written as $S_{out}(t) = (A(t) - I_{PFC \rightarrow AMYG}) \cdot g(\Delta t)$.

Before the transition to the NMG architecture design, where the proposed framework is described, Table~\ref{tab3} maps the sensorimotor system components to the microgrid nodes engaged in the adaptive response regulation scheme under precursor-disturbance events.

\begin{table*}[!htbp]
\caption{Sensorimotor Circuitry – Microgrid Architecture Mapping}
\label{tab3}
\centering
\begin{tabularx}{\textwidth}{|Y|Y|Y|Y|Y|}
\hline
\textbf{Sensorimotor Circuitry} & \textbf{Role in Gating} & \textbf{Neuro-Microgrid Architecture Analog} & \textbf{Role in Control/Protection}\\
\hline
Thalamus (Th) & Fast relay of sensory inputs & PMUs/Telemetry Bus & Capture raw electrical signals (voltage, current, frequency)\\
\hline
Sensory Cortex (SC) & Refined processing, contextual representation & Analytics/Feature Extraction & Feature extraction, distinguish nuisance vs. precursor\\
\hline
Amygdala (AMYG) & Fast excitatory drive (reflexive response) & Reflex Protection Controller (fast inverter/breaker) & Immediate trip/protection tendency\\
\hline
Prefrontal Cortex (PFC) & Inhibitory/facilitatory modulation (contextual) & EMS/Supervisory Controller & Classify disturbances, supervises reflex actions, suppresses false trips, or primes adaptive action\\
\hline
Striatum/Basal Ganglia & Gate integration, apply timing-dependent function & Gate Logic Module & Apply gating factor to scale protection drive\\
\hline
Pontine Reticular Nucleus (PnC) & Motor gateway for startle reflex & Protection/Actuation Gateway (breakers, FACTS, ESS) & Physical enforcement of gated control signals\\
\hline
Motor Output & Final motor response (blink, body jerk) & System Action & Executed control/protection outcome (trip, support, curtail)\\
\hline
\end{tabularx}
\end{table*}

\section{Hierarchical Neuro-Microgrid Gating Framework}

\subsection{Sensorimotor Gating-inspired Microgrid Control}

We refer to the proposed Hierarchical Neuro-Microgrid control scheme as Sensorimotor Gating-Inspired Neuro-Microgrid (SG-NMG) architecture. The SG-NMG builds upon the parallels between brain circuits of PPI and PPF and the HCA of microgrids \cite{li2023hierarchical, vasquez2010hierarchical}. As already mentioned, in the brain, sensory information is rapidly relayed via the Thalamus, processed in the SC, evaluated by the Amygdala (fast excitatory drive), and modulated by the PFC through inhibitory or facilitatory control. Then, the Striatum/Basal Ganglia apply a gating function before signals reach the PnC, producing the final motor output. Analogously, in Neuro-Microgrids, Phasor Measurement Units (PMUs) act as the sensory interface, Analytics Engines refine event classification, Reflex Protection Controllers provide fast protective actions, and the Energy Management System (EMS) exerts supervisory inhibition (no rapid corrective actions 
are needed) or facilitation (intensive corrective actions are needed). A Gate Logic Module then integrates these signals before triggering the Actuation Layer (e.g., breakers, Flexible AC Transmission Systems (FACTS), or ESS).

\subsubsection{Primary Control (PPI as protection damping)}

At the primary level, the SG-NMG emulates the reflexive behavior of the Amygdala. Local controllers (droop control in DGs, embedded ESS controllers) respond within milliseconds to stabilize voltage and frequency. Here, the role of PPI is prominent, which is to ensure that small precursor disturbances (minor sags, harmonics, or switching spikes) are filtered, preventing nuisance trips or excessive actuation. This corresponds to the suppression of irrelevant stimuli in the brain, ensuring stability without overreaction.

\subsubsection{Secondary Control (balanced adaptation)}

The secondary layer mirrors the limbic loop of refined response \cite{saavedra2024brain}. Through supervisory functions in the EMS, contextual information (e.g., persistence of a disturbance, network topology) is integrated with reinforcement or predictive signals from historical events. If a disturbance is benign, inhibitory pathways dominate, damping protective actions. If the precursor is predictive of a major event, facilitation is engaged, increasing responsiveness. Thus, secondary control incorporates timing-dependent gating (analogous to the PFC–Striatum loop) to adaptively restore nominal voltage and frequency while conditioning the degree of intervention.

\subsubsection{Tertiary Control (PPF as adaptive amplification)}

At the tertiary level, the SG-NMG can employ PPF as an adaptive optimization mechanism. Forecasts of renewable generation, load demand, and market prices can act as "prepulses" that prime the EMS. Rather than filtering, the system amplifies sensitivity to these predictive signals, proactively adjusting schedules, dispatching ESS, or curtailing loads. This resembles PPF in the brain, where preparatory cues enhance responsiveness to salient events. In this way, tertiary control strategically aligns economic objectives with resilience, enabling the microgrid to participate in markets while safeguarding critical operations.

\subsection{Generalized SG-NMG Architecture and Workflow}

\begin{figure*}[!htbp]
\centering
\includegraphics[trim={0cm 0cm 0cm 0cm},clip,width=1\textwidth]{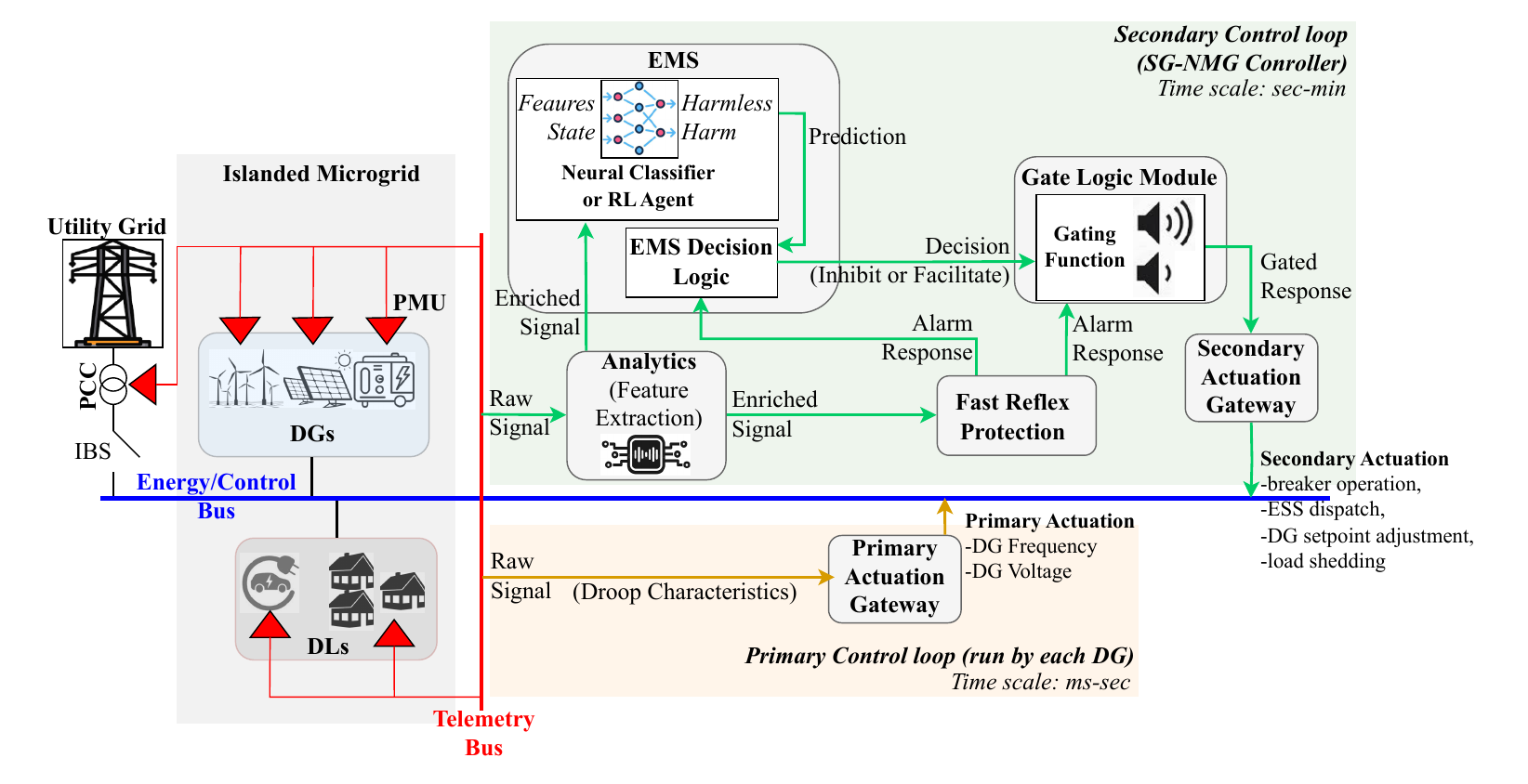}
\caption{Generalized Hierarchical SG-NMG Controller Architecture for Primary and Secondary Control. DG: Distributed Generator; DL: Distributed Load; PMU: Phasor Measurement Unit; SG-NMG: Sensorimotor Gating-based Neuro-Microgrid.}
\label{fig:fig5}
\end{figure*}

The architecture and workflow of the proposed SG-NMG can be described as shown in Fig.~\ref{fig:fig5}. A precursor disturbance is first detected by PMUs and sensors (corresponding to the thalamic role in the brain). This raw signal is then used for feature extraction by Analytics modules (analogous to the sensory cortex). These extracted features will be later used to determine whether the event is a harmless fluctuation or a predictor of a major fault. In practice, features such as the magnitude, duration, frequency content, and rate of change of voltage and current signals are extracted from the raw measurements. For example, a short-duration voltage sag with low harmonic distortion may be later classified as a benign transient caused by capacitor switching, while a sustained sag with high harmonic distortion or rapid current rise may indicate the onset of a fault.

Reflex Protection Controllers (corresponding to AMYG) generate a fast protective drive as an immediate response, reflecting that the system enters into an alarming control state. At the same time, the EMS (corresponding to the PFC) classifies the input signal (and its features) to modulate the reflex drive. The classification can be achieved through a combination of signal processing and machine learning techniques. Data-driven classifiers, such as support vector machines, neural networks, or reinforcement learning agents, can be trained on historical event/feature data to discriminate between "harmless" or "major fault" classes, adapting properly to different grid conditions. In this way, the analytics modules act as the sensory cortex of the microgrid, transforming raw stimuli into enriched representations that allow the supervisory controller (EMS) to decide whether to inhibit or facilitate a protective response. If the precursor is assessed as irrelevant, inhibition is applied (corresponding to PPI) and the protective action is suppressed. If the precursor is predictive of a major disturbance, facilitation is applied (corresponding to PPF), which enhances the responsiveness of the system. The Gate Logic Module (analogous to the striatum and basal ganglia) integrates these signals and applies a conditional gating factor that up- or down-scales the final control action. Finally, the Protection and Actuation Layer (corresponding to the PnC) executes the gated command through breakers, FACTS, or ESS, resulting in the actual system output such as tripping, damping, or reinforcement.

An illustrative scenario can clarify this process. Consider a microgrid operating in islanded mode when a small voltage sag occurs. Sensors and analytics identify whether this sag is simply a benign transient, such as capacitor switching, or an early indicator of a fault. If the sag is benign, the EMS suggests inhibition and the reflexive protection is suppressed, avoiding an unnecessary trip. If the sag is predictive of a line fault, the EMS suggests facilitation which boosts the reflexive controller and triggers early actions such as pre-activating ESS injection. When the fault materializes, breakers and load shedding operate more effectively, ensuring stability is maintained. In this example, PPI functions as protective damping that suppresses overreaction, while PPF functions as adaptive amplification that primes resources for a stronger and proactive response.

To concretely describe the workflow in Fig.~\ref{fig:fig5} considering both primary and secondary control loops (tertiary loop is omitted for simplicity), the following steps are identified:

\textbf{Step 1 - Event detection (Telemetry Bus; ms).} A small disturbance (e.g., sag, harmonic burst) appears on the Energy/Control Bus. PMUs and local meters stream raw signals on the Telemetry Bus.

\textbf{Step 2 - Immediate inner loop engagement (Primary control; ms–s).}
Raw local measurements feed the Primary Actuation Gateway (droop characteristics at each DG). Inner current/voltage loops and droop laws produce fast stabilizing actions (e.g., P and Q adjustments). This loop is self-contained and always closed, ensuring basic voltage/frequency stability even if higher layers are unavailable.

\textbf{Step 3 - Feature extraction (Analytics; ms–s).}
In parallel, the raw signals enter Analytics for Feature Extraction. Time/frequency-domain descriptors can be computed (e.g., timing events, magnitude, duration, Rate of Change of Frequency (RoCoF), Total Harmonic Distortion (THD)). Analytics outputs an enriched signal to both the EMS and the Fast Reflex Protection block.

\textbf{Step 4 - Fast reflex appraisal (Fast Reflex Protection; ms–s).}
Using the enriched signal and local limits, Fast Reflex Protection evaluates instantaneous/definite-time criteria and generates an alarm response when a potentially dangerous trend is detected (reflex tendency to trip, shed, or inject Volt-Amperes reactively (VAR)).

\textbf{Step 5 - Supervisory classification and decision (EMS; s–min).}
The EMS hosts the  trained classifier or RL agent. It ingests the enriched features and (optionally) the alarm response, and outputs a prediction (harmless precursor vs. harmful precursor) with confidence. The EMS Decision Logic converts this into a decision: inhibit (suppress reflex), facilitate (pre-boost response), or (optionally) neutral.

\textbf{Step 6 - Conditional gating (Gate Logic Module; s–min).}
The Gate Logic computes a gating function $g(\cdot)$ that depends on context indicators like persistence, prediction by the EMS model, clustering of precursors, and topology state. It scales the EMS decision into a gated response, i.e. $g(\cdot)<1$ for PPI-like inhibition (protective damping), or $g(\cdot)>1$ for PPF-like facilitation (adaptive amplification), or $g(\cdot)=1$ for neutral pass-through.

\textbf{Step 7 - Secondary execution (Secondary Actuation Gateway; s–min).}
The gated response is translated into secondary actions that bias but do not break the inner loop: (i) inhibition to raise trip thresholds/delays, desensitize shedding, hold ESS, or (ii) facilitation to pre-arm breakers, lower protection delays, pre-dispatch ESS, adjust DG set-point offsets $\Delta f$, $\Delta V$. These offsets enter summing nodes ahead of droop references (outer path), keeping the primary loop independent.

\textbf{Step 8 - System-level effect (Energy/Control Bus).}
Both primary (fast) and secondary (gated) actions materialize on the Energy/Control Bus via the Actuation Gateways connecting breaker states, ESS power/VAR injection, DG set-point adjustments, and selective load shedding. The result is either suppression of nuisance events (PPI) or proactive strengthening against imminent faults (PPF).

\textbf{Step 9 - Post-event reinforcement (learning).}
The EMS quantifies the outcomes (frequency deviation area, overshoot, RoCoF limits hit, false/missed trips, ESS stress, service continuity metrics). These serve as reinforcement signals to retrain the classifier or update the RL policy, and to recalibrate gating parameters. Over time, the EMS learns when to inhibit or facilitate and how strongly to gate.

\textbf{Step 10 - Fail-safe hierarchy.}
If the EMS/Gate Logic path is unavailable, the primary closed loop still stabilizes Voltage/Frequency, and Fast Reflex Protection enforces hard safety limits. The outer path only adds conditional inhibition/facilitation, without compromising inner-loop stability.

\subsection{Supporting RL-based Feedback Loop in SG-NMG}

An RL agent in this setting is hosted in EMS and can be trained by interacting with the microgrid environment, either through simulations or digital twin platforms. At each decision step, the agent receives the extracted features as the state, selects an action (inhibit, facilitate, or allow normal operation), and observes the outcome of the system. If the action leads to stable operation without unnecessary trips, the agent receives a positive reward. If the action leads to instability, false tripping, or failure to respond to a real fault, the agent is penalized. By iteratively updating its policy using reinforcement signals, the agent gradually learns to classify precursors more accurately and to apply the appropriate inhibitory or facilitatory modulation. In this way, the EMS evolves into an adaptive supervisory module that mirrors the brain’s ability to refine gating decisions through experience.

\section{Challenges and Research Opportunities}

Translating sensorimotor gating concepts into microgrid control raises several open challenges. A central issue is the mathematical modeling of gating functions for power systems. Unlike traditional droop or PI controllers, gating decisions in NMGs cannot always rely on a simple timing interval between a prepulse and a pulse, as in biological systems. Instead, they must be conditional on feature patterns of precursor–disturbance events, such as magnitude, rate of change, persistence, or spectral content. Developing models that capture how the EMS uses these enriched features to decide whether to inhibit or facilitate responses requires multiple tests for different modeling selections. Another challenge lies in the coordination of gating across multiple timescales. Events in microgrids unfold from fast transients (milliseconds) to slower disturbances (seconds to minutes), and determining at which layer and under what conditions gating should be applied requires new strategies for multi-layer adaptation.

Equally important are the challenges of practical implementation and validation. Realizing gating functions in distributed and decentralized architectures demands lightweight coordination mechanisms that avoid over-reliance on centralized EMS decisions. This creates opportunities for integrating AI/ML methods, such as RL agents that learn gating policies from experience, neuro-fuzzy controllers that embed expert knowledge, or graph-based models that capture interdependencies across distributed units. Finally, robust validation will require hardware-in-the-loop (HIL) and real-time simulation frameworks to test gating-inspired architectures under realistic dynamics and disturbances. These steps will be essential to move from conceptual analogies toward deployable self-healing and adaptive NMGs.

\section{Future Trends}

Together with the existing brain-inspired microgrid designs, the proposed SG-NMG concept opens several promising directions for future research. One key trend is the evolution toward multi-level adaptive control architectures that explicitly mimic the brain's gating functions. By embedding PPI-like inhibition and PPF-like facilitation across primary, secondary, or even tertiary layers, future microgrids could dynamically balance stability and flexibility under uncertainty. Such approaches will move beyond conventional rule-based protection to create neuro-inspired controllers capable of selectively filtering nuisance events while proactively amplifying responses to credible threats.

Another emerging direction is the integration of digital twins for design, training, and validation. Digital twins of microgrids provide a safe environment to model gating-inspired frameworks, test RL agents, and calibrate gating functions $g(\cdot)$ under diverse scenarios of renewable variability, load fluctuations, and fault disturbances. These platforms enable continuous refinement of supervisory logic and reduce the risks of deploying experimental neuro-inspired mechanisms directly in physical systems.

Looking ahead, the development of NMGs will require cross-disciplinary collaborations that connect insights from neuroscience, control engineering, and power systems. By translating knowledge of neural gating into control algorithms, research in industrial electronics can unlock applications in self-healing grids, cyber-physical resilience, and renewable integration, where proactive adaptation and disturbance filtering are critical. Such convergence promises not only to enhance technical resilience and efficiency but also to adopt a new paradigm in how we design intelligent and adaptive infrastructures for future energy systems.

\section{Conclusion}

This article introduced sensorimotor gating as a powerful parallel for the design of resilient and adaptive microgrid controllers. By drawing inspiration from the brain’s mechanisms (PPI and PPF), we outlined how disturbances in power systems can be selectively suppressed or proactively amplified depending on their features and context. The resulting Neuro-Microgrid paradigm may represent a promising direction for the evolution of smart energy systems. Embedding gating principles across the hierarchical layers of control enables microgrids to behave as self-protective and self-adaptive entities, capable of operating under uncertain environments and optimizing long-term operation. Building on the continuous coupling of neuroscience and industrial electronics engineering, this paradigm may further contribute to the design of future brain-inspired microgrids.

\bibliographystyle{IEEEtran}

\bibliography{bibliography}

\end{document}